\documentclass[aps,prb,twocolumn,groupedaddress]{revtex4-1}

\usepackage{graphicx}
\usepackage[space]{grffile}
\usepackage{dcolumn}
\usepackage{bm}
\usepackage{subfigure}
\usepackage{mathrsfs}
\usepackage{booktabs}
\usepackage{notoccite}
\usepackage{float}
\usepackage{placeins}
\usepackage{enumerate}
\usepackage{gensymb}
\usepackage{amsmath}
\usepackage{color}
\usepackage{siunitx}

\newcommand{\rom}[1]{\uppercase\expandafter{\romannumeral #1\relax}}

\begin{document}

\title{Voltage-dependent reconstruction of layered Bi$_2$WO$_6$ and Bi$_2$MoO$_6$ photocatalysts \\and its influence on charge separation for water splitting }

\author{Quinn Campbell}
\email{quinn.campbell@psu.edu}
\author{Daniel Fisher}
\author{Ismaila Dabo}
\affiliation{Department of Materials Science and Engineering, Materials Research Institute, and Penn State Institutes of Energy and the Environment, The Pennsylvania State University, University Park, PA 16802, USA}

\begin{abstract}
We study the surface stability of the layered bismuth-oxide Bi$_2$WO$_6$ and Bi$_2$MoO$_6$ photocatalysts, which belong to the series of Aurivillius (Bi$_2$A$_{n-1}$B$_{n}$O$_{3n+3}$) perovskites and have been proposed as efficient visible-light absorbers, due to favorable electronic hybridization induced by the Bi 6s and 6p orbitals. We present a Newton--Raphson optimization of the charge distribution at the semiconductor--solution interface using the self-consistent continuum solvation (SCCS) model to describe the influence of the aqueous environment. Our analysis provides a description of the charged interface under controlled pH and applied voltage, and offers a molecular interpretation of the competing structural and electrical factors that underlie the facet-dependent photocatalytic activity of layered Bi$_2$A$_{n-1}$B$_{n}$O$_{3n+3}$ compounds.
\end{abstract}

\maketitle

\section{Introduction}

Artificial photosynthesis is uniquely positioned to alleviate the energy needs of the world population by converting water, carbon dioxide, and sunlight into fuels\cite{Lewis2007}. An outstanding challenge facing this technology is to develop photocatalysts of low cost and of high durability \cite{McKone2014}. The search for new photocatalysts focuses on metal oxides due to their stability in water and their chemical versatility \cite{Woodhouse2009,Woodhouse2005}. However, few of these oxides can absorb visible light, limiting their use as photoelectrodes. It is thus critical to optimize the band gap of these materials to enable them to operate efficiently under sunlight. 

\begin{figure}
	\includegraphics[width=\columnwidth]{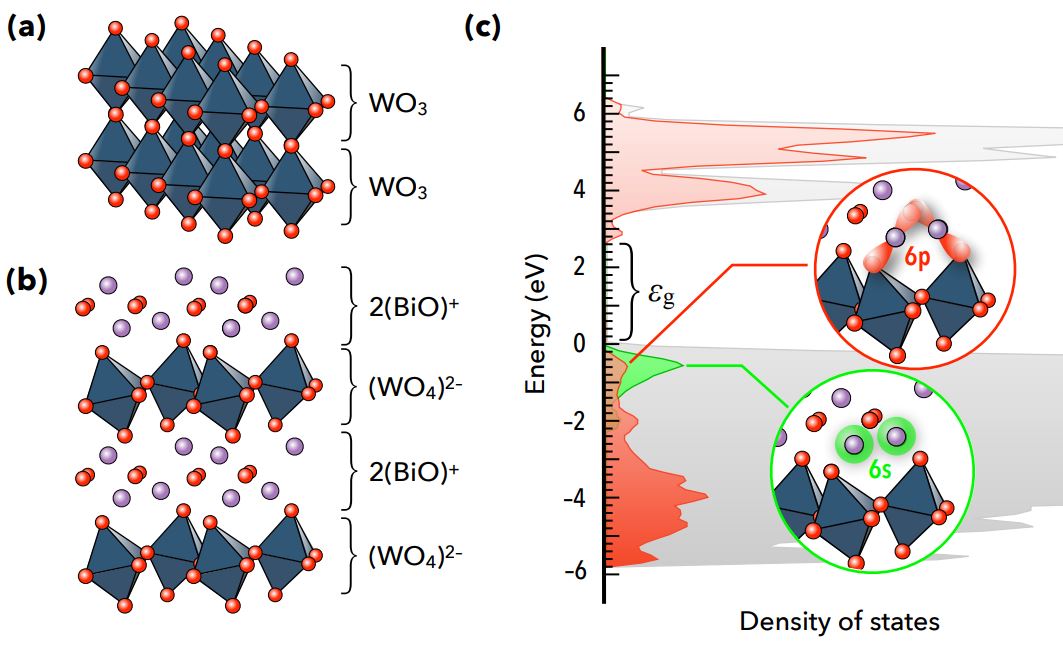}
	\caption{ \small The intercalation of bismuth oxide into the pseudocubic lattice of tungsten oxide WO$_3$ (panel a) generates a layered bismuth-oxide (Aurivillius) Bi$_2$WO$_6$ structure (panel b) that is characterized by strong hybridization of the bismuth 6s and 6p states with the valence bands of the original WO$_3$ oxygen 2p states as seen from the bismuth 6s (green) and 6p (red) projected density of states and from the total (gray) density of states including scissor corrections \cite{Gritsenko1995,Kuisma2010}, dominated by the oxygen 2p orbital (panel c). This hybridization causes the band gap $\epsilon_{\rm g }$ of WO$_3$ to decrease, thereby enhancing its ability to absorb sunlight.}
	\label{fig:unit-cells}
\end{figure}

The fractional substitution of the cationic species that compose the metal oxides is an effective means to tune their band gap\cite{Wang2012,Hu2008,Ling2014,Kim2006}. Nevertheless, this approach introduces compositional disorder that increases the rate of electron--hole recombination. In contrast, the intercalation of functional layers into metal oxides provides an effective method to control the band gap while preserving or enhancing charge separation \cite{Kubacka2012}. An example of the effectiveness of this method to modify the electronic structure of WO$_3$ is show in Fig.~\ref{fig:unit-cells}; the intercalation of bismuth oxide layer in WO$_3$ creates electronic states above the valence band maximum, thereby reducing the band gap of WO$_3$. In addition to WO$_3$, this method has been successfully used to enhance light absorption in a range of semiconducting ABO$_3$ and BO$_3$ metal oxides for use as photocatalytic electrode materials\cite{Kim2004,Thakral2010,Saison2015,Amano2008,Amano2009,Shimodaira2006,Ng2012,Zhang2011}.

The ability to design metal oxides that are compatible with the solar spectrum by intercalating functional oxide layers provides a strong motivation to further study the photocatalytic activity of this family of layered semiconductors. In particular, the accurate determination of the photocatalytic mechanisms that take place on the surface of the electrode requires one to know the interfacial structure under applied voltage and controlled pH.

In this work, we address critical questions surrounding the surface termination of Bi$_2$WO$_6$ and Bi$_2$MoO$_6$, two prototypical layered oxides of the Aurivillius series, having the generic chemical formula Bi$_2$A$_{n-1}$B$_{n}$O$_{3n+3}$. Since different surface terminations exhibit different catalytic properties, knowing which terminations are most stable under varying environmental conditions is critical to narrowing down the choice of candidate photocatalysts and guiding their synthesis. Experimental studies by Saison {\it et al.} and Zhang {\it et al.} have shown that the (010) crystalline facet dominates the surface structure of Bi$_2$WO$_6$ and Bi$_2$MoO$_6$ particles in electrolytic media \cite{Saison2011a,Zhang2010}. The (010) facet can exhibit multiple terminations, however, ending with either W/Mo tetrahedrons or Bi$_2$O$_2$ layers. While Zhang {\it et al.} have demonstrated that the stability of the (010) facet is strongly enhanced by the adsorption of oxygen species, the specific molecular arrangement of the oxygen-terminated surface is still poorly understood, precluding the analysis of photocatalytic trends as a function of surface structure. 

First-principles modeling provides a powerful approach to investigating the surface stability and electrical response of solvated semiconductor electrodes with atomic-level precision and quantum-mechanical accuracy \cite{Kharche2014,Pham2017a,Selcuk2016,Blumenthal2017}. However, the application of this approach to layered photoelectrodes faces two major problems. The first is accessing the equilibrium charge--voltage response of their various surface terminations, and the second is assessing the surface free energy of different layers within the electrode.

Regarding the first problem, predicting the charge--voltage response of a semiconductor electrode entails describing the accumulation of charge at its surface and within the subsurface depletion region. Recently, we have developed a quantum--continuum model that incorporates an electronic-structure Kohn--Sham treatment of the surface region with a semiclassical Mott--Schottky representaion of the depletion layer to provide a complete description of the electrification of the interface \cite{Campbell2017}. Nevertheless, the application of this quantum--continuum model requires the user to perform a number of self-consistent calculations to optimize the charge distribution between the surface and the bulk of the photoelectrode. Here, we circumvent this step by developing a  robust, fully automated algorithm that directly converges to the optimal distribution. To solve the second problem, we extend previous methodologies to find the free energy of each ionic layer as a function of potential and pH, allowing us to consistently calculate the surface free energy of each layer of each termination.

The paper is organized as follows. We present the electronic-structure computational procedure in Sec.~\ref{sec:comp-det} and describe the finite-difference Newton--Raphson algorithm for simulating semiconductor electrodes under applied voltage in Sec.~\ref{sec:descent}. We then generalize this approach in Sec.~\ref{sec:surf-e-adsorbates} to predict the coverage-dependent surface free energy of layered semiconductors. Finally, Sec.~\ref{sec:results} reports our computational results on the surface stability and junction characteristics of several Bi$_2$WO$_6$ and Bi$_2$MoO$_6$ terminations along the (100) and (010) orientations, with a focus on understanding the structural and electronic evolution of the surface during a potential sweep.

\section{Methods}
\subsection{Embedded electronic-structure calculations}
\label{sec:comp-det}
\begin{figure}
	\includegraphics[width=\columnwidth]{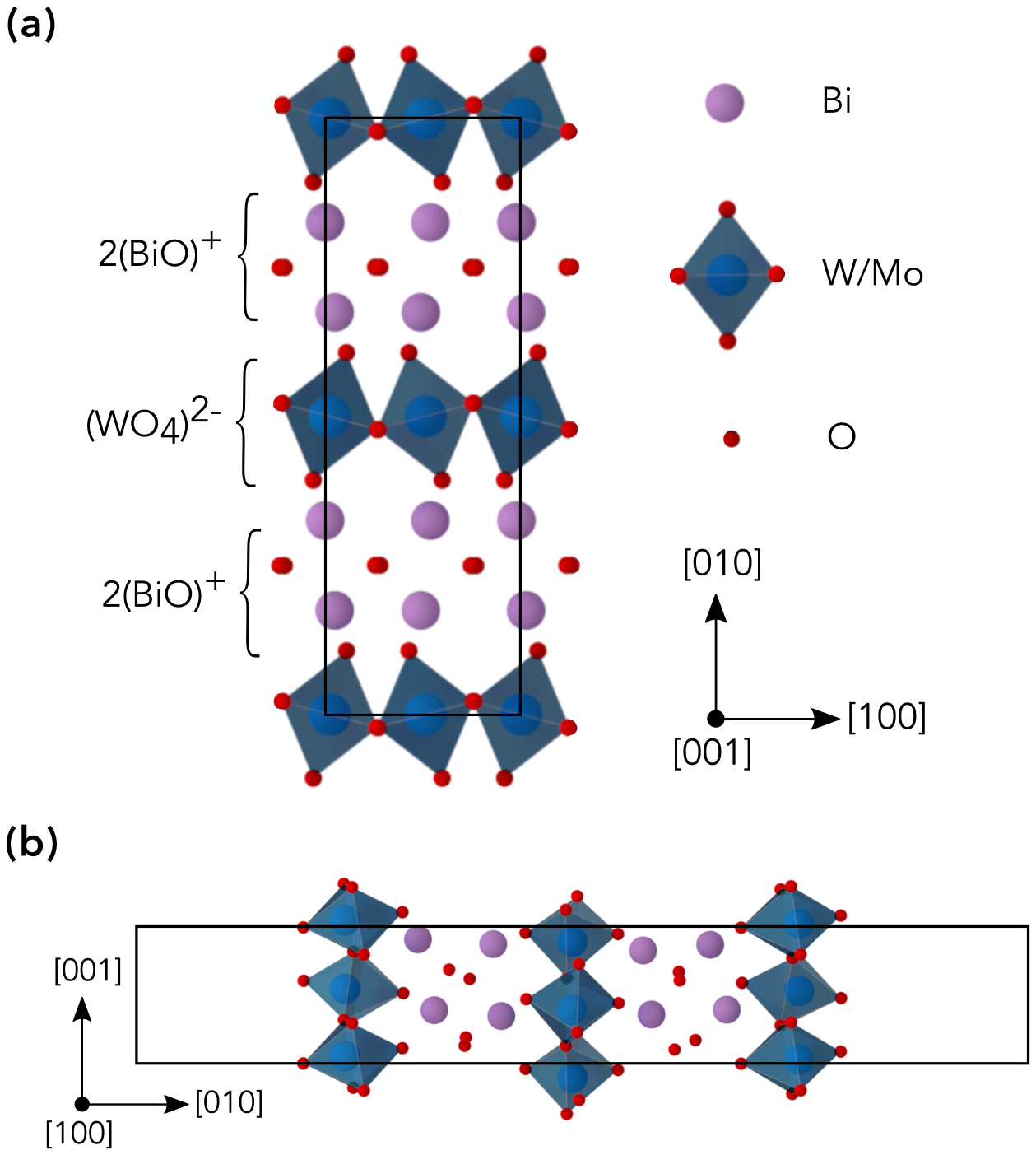}
	\caption{ \small (a) Layered structure of bulk Bi$_2$WO$_6$ and Bi$_2$MoO$_6$ compounds, showing the alternating tilt of the W and Mo octahedra upon geometry optimization. (b) Slab structure used for supercell surface calculations.}
	\label{fig:bulk-structure}
\end{figure}

To examine the structural properties of Bi$_2$WO$_6$ and Bi$_2$MoO$_6$, we first perform an optimization of their bulk crystalline geometry. We use the {\sc pw} implementation of density-functional theory (DFT) within the Quantum-Espresso distribution for materials simulation \cite{Giannozzi2009}. We employ the Perdew--Burke--Ernzerhof exchange-correlation functional \cite{Perdew1996} with pseudized atomic cores from the SSSP repository \cite{Lejaeghere2016}, which provides an extensively validated library of pseudopotentials. We use projector augmented wavefunction (PAW) descriptions of each ionic core. We sample the Brillioun zone with a shifted 4 $\times$ 2 $\times$ 1 Monkhorst--Pack grid and 0.03 Ry of Marzari--Vanderbilt smearing \cite{Marzari1997}. We select wavefunction and charge density kinetic energy cutoffs of 150 Ry and 600 Ry, respectively. The resulting optimized bulk geometries are shown in Fig.~\ref{fig:bulk-structure}a. The calculated lattice parameters are $a$ = 5.56 \AA\, $b$ = 16.84 \AA\ , $c$ = 5.59 \AA\ for Bi$_2$WO$_6$ and $a$ = 5.66 \AA\ , $b$ = 16.53 \AA\ , $c$ = 5.67 \AA\ for Bi$_2$MoO$_6$, in close agreement with experimental data \cite{Yanovskii1986}.
\begin{figure*}
	\includegraphics[width=\textwidth]{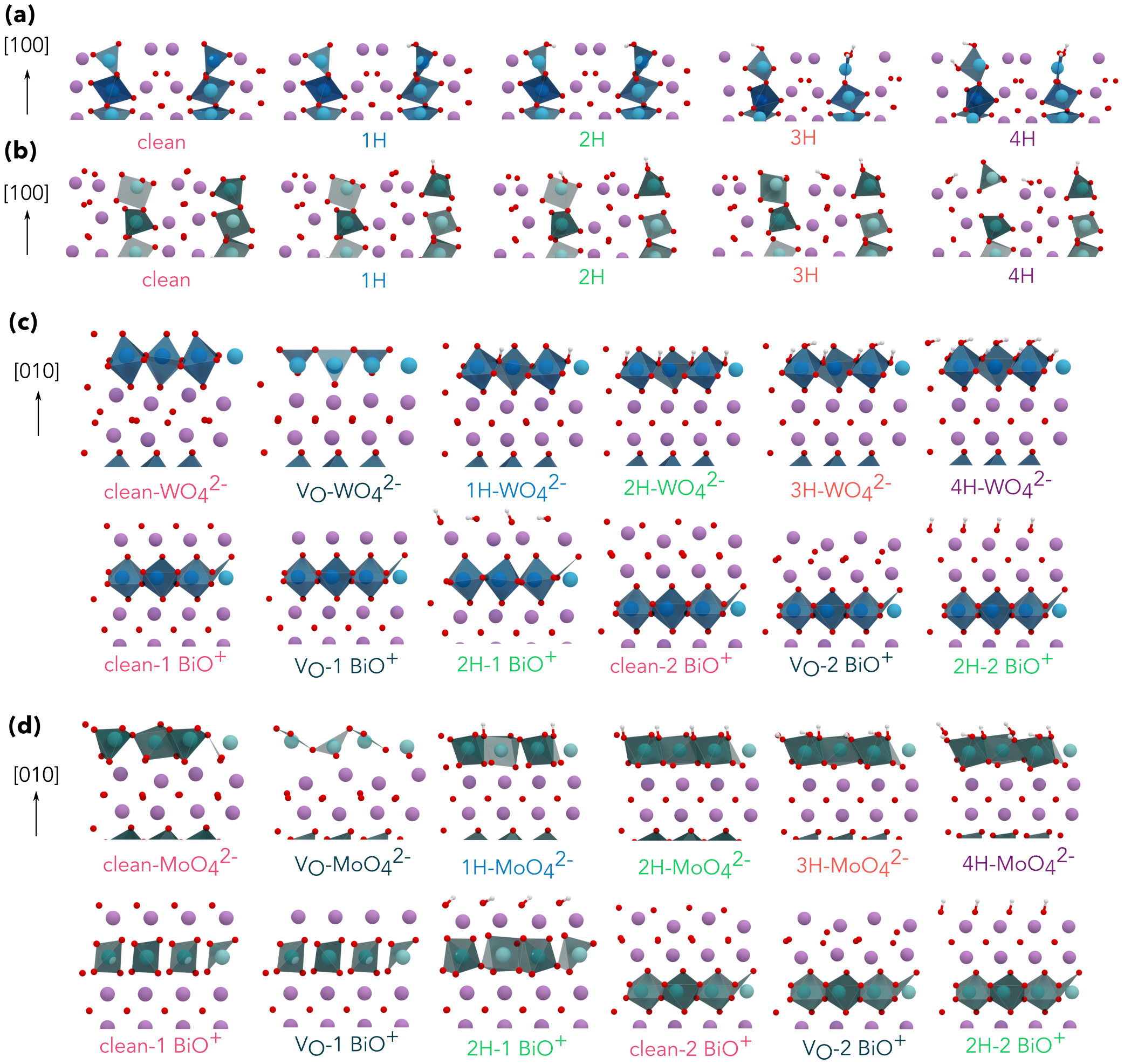}
	\caption{ \small Surface terminations for (a) Bi$_2$WO$_6$ (100), (b) Bi$_2$MoO$_6$ (100), (c) Bi$_2$WO$_6$ (010), and (d) Bi$_2$MoO$_6$ (010). The notation {\it n}H indicates {\it n} hydrogen adsorbed per unit cell. Since the (010) facet can be terminated with WO$_4^{2-}$ layers, one or two BiO$^+$ layers. V$_{\rm O}$ indicates oxygen vacancies on the surface.}
	\label{fig:structures}
\end{figure*}

To determine the voltage-dependent surface restructuring of Bi$_2$WO$_6$ and Bi$_2$MoO$_6$, we then create  symmetric slab structures with the form shown in Fig.~\ref{fig:bulk-structure}b and the surface terminations shown in Fig.~\ref{fig:structures}. Specifically, for the (010) facet, we tested W/Mo terminated slabs with both one and two layers of Bi termination. For the (100) surface we just considered a unique termination, since the (100) crystalline orientation does not exhibit an alternating layer pattern. We center the slab in the supercell with a vacuum height of 14 \AA\ .  In all cases, a slab thickness of five layers is found to be sufficient to achieve a convergence of 50 meV for the Fermi energy and of 60 meV per unit cell for surface energies as shown in the supporting information. 

We employ the {\sc environ} module, which implements a self-consistent continuum solvation (SCCS) model to describe the implicit immersion of the quantum system in aqueous media \cite{Andreussi2012}. Dielectric cavities are introduced around each lateral facet of the slab with local dielectric permittivity written on the semiconductor side as $\epsilon (\bm{r}) =\exp[(\zeta(\bm{r}) -\sin(2\pi\zeta(\bm{r}))/2\pi)\ln\epsilon_{\rm sc}]$ where $\epsilon_{\rm sc}$ is the dielectric constant of the semiconductor; a similar expression can be taken for the solution with $\epsilon_{\rm sc}$ replaced by the dielectric constant of the medium $\epsilon_{\rm m}$. Here, $\zeta(\bm{r}) =(\ln \rho_{\rm max} - \ln\rho(\bm{r}))/(\ln \rho_{\rm max} -\ln \rho_{\rm min}) $ is a smooth switching function, marking the gradual dielectric transition between the quantum and continuum regions based on the charge density of the electrode, where $\rho_{\rm min}$ and $\rho_{\rm max}$ serve as the density thresholds specifying the inner and outer isocontours of the dielectric cavity. The SCCS model also includes contributions from the external pressure, solvent surface tension, and solvent dispersion and repulsion effects. The surface tension is described by $G_{\rm cav}=\gamma S$ and the dispersion and repulsion effects by $G_{\rm dis+rep} = \alpha S + \beta V $. Here, $\gamma$ is the solvent surface tension, taken from experiment, $\alpha$ and $\beta$ are fitted parameters, and $S$ and $V$ are the quantum surface and volume of the solute, defined as $S = \int d\mathbf{r}(d\Theta/d\rho)  |\nabla \rho|$ and $V=\int d\mathbf{r} \Theta(\rho) $, where $\Theta$ is another smooth switching function, defined by $\Theta(\rho) = (\epsilon_{\rm s} - \epsilon(\rho))/(\epsilon_{\rm s}-1)$. We utilize the parameterization of Andreussi {\it et al.}, where $\epsilon_{\rm m} = 78.3$ is the dielectric constant of the water, $\rho_{\rm max}= 5 \times 10^{-3}$ a.u., $\rho_{\rm min}= 1 \times 10^{-4}$ a.u., $\gamma = 72.0$ dyn/cm, and $\alpha = -22$ dyn/cm \cite{Andreussi2012}. There has recently been some discussion that the introduction of the volume term is unphysical for slab surfaces, introducing an energy dependence on the overall size of the slab \cite{Fisicaro2017,Andreussi2018}. To clarify the impact of the volume parameter, we set $\beta = 0$ and compared final surface stability results. The elimination of the volume term led to minor changes in the final energy reported for each slab. Since the volume of all the slabs tested was roughly the same, however, the change in energy was essentially constant across all surfaces tested, leading to no alteration in the final reported surface stability. We use a dielectric constant for the semiconductor of $\epsilon_{\rm sc} = 5.7$, found from linear perturbation calculations \cite{Baroni2001}.
\subsection{Newton--Raphson charge optimization}
\label{sec:descent}
In order to simulate the electrified semiconductor--solution interface, including the contributions from surface states to the charge--voltage response of the electrode, we employ the computational approach described in Ref.~\onlinecite{Campbell2017}. In this model, we embed a quantum-mechanical description a semiconductor electrode surface between a Poisson-Boltzmann distribution of ionic charges and a Mott--Schottky distribution of charged defects on the electrolyte and semiconductor sides, respectively. We then impose that the Fermi energy be constant across the entire interface. To enforce this condition, we must determine the equilibrium amount of charge within the explicit surface region and the implicit bulk depletion region. For a detailed description of the computational procedure, we refer the reader to the supporting information.

While our previously proposed computational approach solved the problem of aligning the Fermi level with the correct charge distribution manually, here we develop a numerical approach to iteratively optimize the charge on the explicit quantum-mechanical part of the system when the total charge on the electrode $q$ is given. The charge $q^{\rm surf}$ is updated at each iteration $n$ using the Newton--Raphson algorithm
\begin{equation}
	q_{n +1}^{\rm surf}= q_{n}^{\rm surf} - (\Delta E_{\rm F})_{n}/ (\Delta E_{\rm F})'_{n},
	\label{eq:newton-raph}
\end{equation}
where $(\Delta E_{\rm F})_n$ is the difference between the bulk and surface Fermi levels at iteration $n$, defined as
\begin{equation}
	(\Delta E_{\rm F})_n= E_{\rm F}^{\rm bulk} - E_{\rm F}^{\rm surf}.
	\label{eq:derivative}	
\end{equation} 
The derivative $(\Delta E_{\rm F})'_n$ with respect to the explicit charge of the quantum mechanical region is evaluated  using the finite-difference equation
\begin{equation}
	(\Delta E_{\rm F})'_{n} = ((\Delta E_{\rm F})_{n}  -  (\Delta E_{\rm F})_{n-1})/(q_{n}^{\rm surf}  -  q_{n-1}^{\rm surf}).
\end{equation}
Equation \ref{eq:newton-raph} leads to a smooth convergence of the charge starting from a reasonable estimate of the fraction of charge that is located in the explicit interface region. In specific terms for the Aurivillius compounds, we used $q^{\rm surf} = 0.7 q$ as the initial condition with the frontier between the explicit and implicit of the semiconductor located two layers within the electrode and a dopant concentration of $10^{18}$ cm$^{-3}$ for both facets.

This method operates similarly to a structural optimization, involving a series of self-consistent field iterations. Convergence of surface charge within 1\% of the total electrode charge typically requires ten Newton--Raphson steps. We developed this code within Quantum-Espresso 6.1, and the {\sc environ} 0.2 module \footnote{Our implementation of a Newton-Raphson solver for the equilibrium charge--voltage behavior of a semiconductor can be found at https://github.com/quantumquinn/qe-environ-sc.}.

\subsection{Voltage- and pH-dependent stability}
\label{sec:surf-e-adsorbates}
Calculating surface stability as a function of potential and pH also necessitates including (1) the chemical potential of the adsorbing species and (2) the chemical potential of the injected electronic charge in the evaluation of the surface energy.

The free energy of a surface with $N_{\rm Bi}$ bismuth layers, $N_{\rm W}$ ($N_{\rm Mo}$) tungsten (molybdenum) oxide layers, $N_{\rm H}$ hydrogen adsorbates, $N_{\rm O}$ oxygen adsorbates, and a charge $q$ can be expressed as 
\begin{multline}
\Delta G(N_{\rm Bi},N_{\rm W},N_{\rm H},N_{\rm O},q) = \\ \Delta G(N_{\rm Bi},N_{\rm W},N_{\rm H},N_{\rm O},q=0) + \int_{0}^{q} \Phi(q') dq', 
\label{eq:surface-energy-vs-charge}
\end{multline}
where $\Phi(q)$ is the charge-dependent electrical potential of the interface. To calculate $\Delta G(N_{\rm Bi},N_{\rm W},N_{\rm H},N_{\rm O},q=0)$ we evaluate the total energy of this structure, $E(N_{\rm Bi},N_{\rm W},N_{\rm H},N_{\rm O},q=0)$, and subtract the chemical potential of each ionic species $\mu({\rm BiO}^+)$ and $\mu({\rm WO}_4^{2-})$, hydrogen ion $\mu({\rm H}^+)$, and oxygen ion $\mu({\rm O}^{2-})$:
\begin{multline}
\Delta G(N_{\rm Bi},N_{\rm W}N_{\rm H},N_{\rm O},q=0) = \\E(N_{\rm Bi},N_{\rm W},N_{\rm H},N_{\rm O},q=0)  -N_{\rm Bi}(\mu({\rm BiO}^+) - e_0 \Phi) \\-N_{\rm W}(\mu({\rm WO}_4^{2-}) +2 e_0 \Phi)- N_{\rm H}  (\mu({\rm H}^+) - e_0 \Phi) \\ - N_{\rm O} (\mu({\rm O}^{2-}) + 2 e_0 \Phi)
\end{multline}
where $\Phi$ is the electronic potential of the electrode, as previously proposed by Rong and coworkers for calculating surface free energy of oxide compounds \cite{Rong2015a,Rong2016}.

Following the computational hydrogen-electrode method\cite{Man2011a,Norskov2004,Rossmeisl2007}, the energy of hydrogen ions in solution can be determined from 
\begin{equation}
{\rm H}^+ + e^- \leftrightarrow \frac{1}{2} {\rm H}_2(g),
\end{equation}
which is at equilibrium at the potential of the reversible hydrogen electrode. Therefore the equilibrium chemical potential of ${\rm H}^+$ is 
\begin{equation}
\mu^{\circ}({\rm H}^+) = \frac{1}{2} E(\rm{H}_2) + e_0\Phi^{\circ}_{\rm H/H^+},
\label{eq:H+}
\end{equation}
where $E({\rm H}_2)$ is the energy of molecular hydrogen in the gas phase. We can then express the chemical potential of the solvated proton as 
\begin{equation}
\mu({\rm H}^+) = \mu^{\circ}({\rm H}^+) -k_{\rm B}T \ln(10) {\rm pH}.
\label{eq:mu_H+}
\end{equation}
where $k_{\rm B}$ is the Boltzmann constant and $T = 300 K$ is the ambient temperature.
Similarly, the energy of removing an ${\rm OH}^-$ ion out of solution can be evaluated from
\begin{equation}
 {\rm OH}^- + {\rm H}^+ \leftrightarrow {\rm H}_2{\rm O},
\label{eq:OH}
\end{equation}
yielding
\begin{equation}
\mu^{\circ}({\rm OH}^-) = \mu^{\circ}({\rm H}_2{\rm O}) - \mu^{\circ}({\rm H}^+),
\label{eq:mu0_OH}
\end{equation}
where $\mu^{\circ}({\rm H}_2{\rm O})$ is approximated by the total energy of a single water molecule in vacuum. Noting that pOH $=$ 14 $-$ pH, we can find the chemical potential as 
\begin{equation}
\mu({\rm OH}^-) = \mu^{\circ}({\rm OH}^-) -k_{\rm B}T \ln(10)(14- {\rm pH})
\label{mu_OH}
\end{equation}

Finally, the energy of O$^{2-}$ can be obtained from the equilibrium relation
\begin{equation}
\mu({\rm O}^{2-}) = \mu({\rm OH}^-) - \mu({\rm H}^+).
\end{equation}
With these equations in hand, we can determine the  surface adsorbates as a function of potential and pH.

For calculating the terminal layer of the surface at equilibrium, we also need to determine the energy of removing BiO$^+$ and WO$_4^{2-}$ from the surface. To this end, we write the chemical reaction
\begin{equation}
{\rm BiO}^+ + 2{\rm H}^+ + 3 e^-\leftrightarrow {\rm Bi} + {\rm H}_2{\rm O},
\label{eq:BiO}
\end{equation}
which is in equilibrium at $\Phi^{\circ}_{\rm Bi/Bi^+} = 0.320$ V \cite{Bard1985}, and from which we can obtain the equilibrium chemical potential of BiO$^+$ in solution under standard conditions:
\begin{multline}
\mu^{\circ}({\rm BiO}^{+}) = \mu^{\circ}({\rm Bi}) + E({\rm H}_2{\rm O}) - 2 \mu^{\circ}({\rm H}^+) \\ + 3 e_0\Phi^{\circ}_{\rm Bi/Bi^+},
\label{eq:mu0_BiO}
\end{multline}
where $\mu^{\circ}({\rm Bi}) $ is the energy of solid bismuth. We then express the chemical potential of the ideal solution as 
\begin{equation}
	\mu({\rm BiO}^{+}) = \mu^{\circ}({\rm BiO}^{+}) + k_{\rm B}T \ln\left[{\rm BiO}^+\right] ,
\label{eq:mu_BiO}
\end{equation}
where $\left[{\rm BiO}^+\right]$ is the concentration of the ${\rm BiO}^+$ ions. (For simplicity, we will consider an electrolyte saturated in ${\rm BiO}^+$, eliminating the natural logarithm term.)

Similarly, we can calculate the energy of  WO$_4^{2-}$ with the following chemical reaction:
\begin{equation}
{\rm Ag}_2{\rm WO}_4 + 2 e^-\leftrightarrow {\rm Ag} + {\rm WO}_4^{2-},
\label{eq:WO4}
\end{equation}
which is at equilibrium at $\Phi^{\circ}_{\rm W/W^{2-}} = 0.466$ V (essentially the same chemical reaction can be used for ${\rm MoO}_4^{2-}$ at $\Phi^{\circ}_{\rm Mo/Mo^{2-}}= 0.4573$ V)\cite{Bard1985}. We can thus express the energy as
\begin{multline}
\mu^{\circ}({\rm WO}_4^{2-}) = \mu^{\circ}({\rm Ag}_2{\rm WO}_4) - 2 \mu^{\circ}({\rm Ag}) -2 e_0 \Phi^{\circ}_{\rm W/W^{2-}},
\label{eq:mu0_WO4}
\end{multline}
where $\mu^{\circ}({\rm Ag}_2{\rm WO}_4)$ is the energy of solid Ag$_2$WO$_4$, and $\mu^{\circ}({\rm Ag})$ is the energy of solid silver. Finally, we derive the chemical potential as:
\begin{equation}
\mu({\rm WO}_4^{2-}) = \mu^{\circ}({\rm WO}_4^{2-}) + k_{\rm B}T \ln\left[{\rm WO}_4^{2-}\right] 
\label{eq:mu_WO4}
\end{equation}
where $\left[{\rm WO}_4^{2-}\right]$ is the concentration of the ${\rm WO}_4^{2-}$ ion in solution. We will again assume the solution is saturated with ${\rm WO}_4^{2-}$ ions.

With the energy of the solvated ions calculated, we can determine the equilibrium energy of each surface termination and adsorbate across a range of electrochemical conditions. These results are reported in the next section.
\section{Results and Discussion}
\label{sec:results}

\begin{figure}
	\includegraphics[width=\columnwidth]{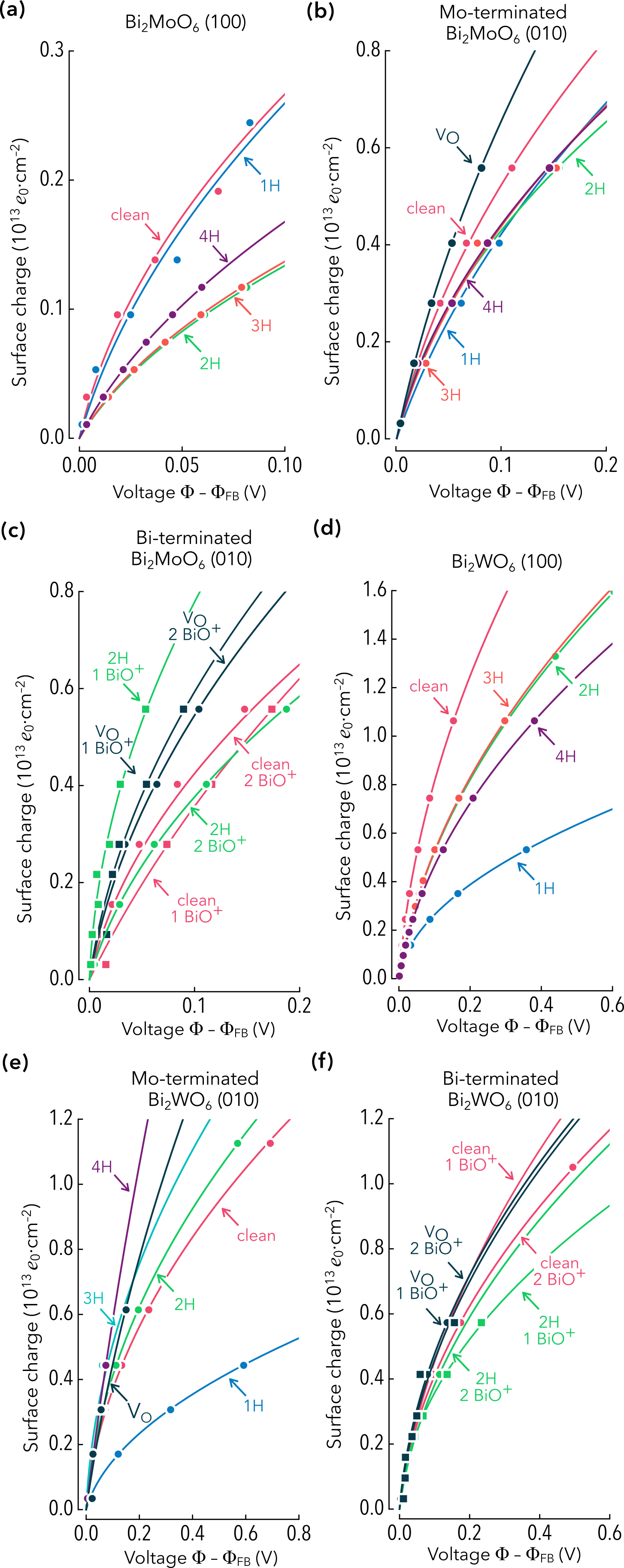}
	\caption{ \small Charge--voltage response of adsorbate-covered  on the Bi$_2$WO$_6$ and Bi$_2$MoO$_6$ photoelectrodes. (a) Bi$_2$MoO$_6$ (100), (b) Mo-terminated Bi$_2$MoO$_6$ (010), (c) Bi-terminated Bi$_2$MoO$_6$ (010), (d) Bi$_2$WO$_6$ (100), (e) W-terminated Bi$_2$WO$_6$ (010), and (f) Bi-terminated Bi$_2$WO$_6$ (010).}
	\label{fig:chg-voltage}
\end{figure}

We determine the electrochemical properties of Bi$_2$MoO$_6$ and Bi$_2$WO$_6$ by calculating the charge--voltage responses of their adsorbate-covered (100) and (010) facets, shown in Fig.~\ref{fig:chg-voltage}. As explained previously, we consider terminations with W/MoO$_4^{2-}$ layers, and one or two layers of BiO$^+$. A first important observation is that the charge profiles of different adsorbates for all but the Bi$_2$WO$_6$ (100) surface are tightly clustered. This trend indicates that charge trapping by surface adsorbates plays a moderate role in the electrical response of Bi$_2$MoO$_6$ and Bi$_2$WO$_6$, and that the specific nature of the adsorbate does not strongly affect the distribution of charge across the interface. 

To confirm and refine these observations, we calculate the Schottky barrier $\Phi_{\rm s}$ of each termination.  The Schottky barrier is the electronic barrier that develops between the bulk of the semiconductor and the surface to compensate the difference between the donor and acceptor levels of the semiconductor and solution respectively. It plays a decisive role in the ability of the interface to conduct photogenerated charge carriers from the bulk semiconductor to the surface and is thus a primary descriptor of the photocatalytic performance of a surface. In the limit of an ideal interface with no charge trapping induced by surface states, the Schottky barrier can be calculated as the difference between the electron-donating and electron-accepting levels on the semiconductor and solution side ($\Phi_{\rm FB}$ and $\Phi$), respectively. However, when adsorption or reconstruction induces surface states, the Schottky barrier height is renormalized by the charge-pinning fraction $\mathscr S$, yielding
\begin{equation}
\Phi_{\rm s} = \mathscr{S} (\Phi - \Phi_{\rm FB}).
\end{equation}
\begin{figure}
	\includegraphics[width=\columnwidth]{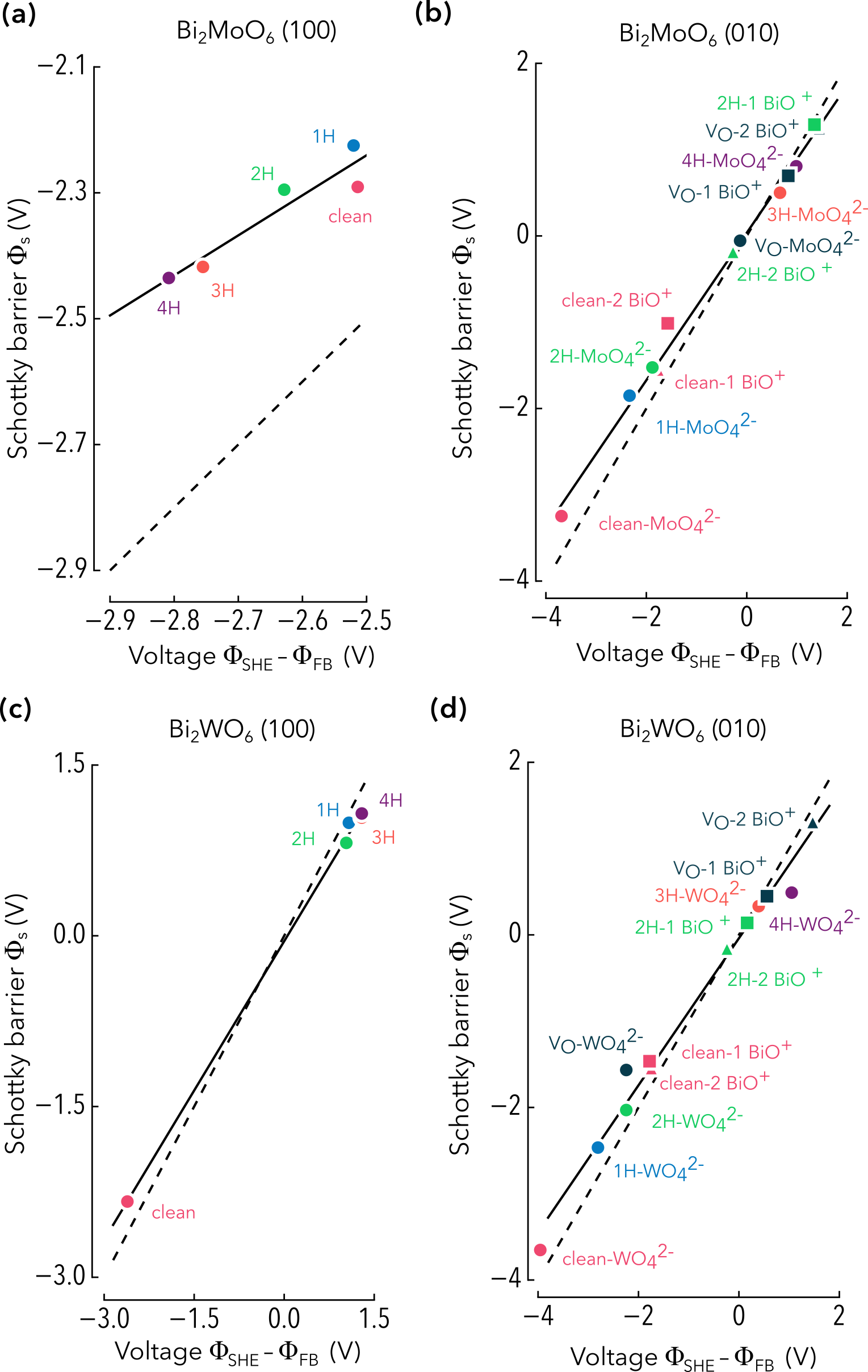}
	\caption{ \small Schottky barriers of different adsorbates and terminal layers at the (a) Bi$_2$MoO$_6$ (100), (b) Bi$_2$MoO$_6$ (010), (c) Bi$_2$WO$_6$ (100), (d) Bi$_2$WO$_6$ (010) surfaces. An ideal semiconductor junction would have a unit slope, $\mathscr{S}=1$ (dashed line). The difference between the dashed line and the line of best fit shows the impact of surface states and adsorbates on lowering the Schottky barrier. The Bi$_2$MoO$_6$ (100), Bi$_2$MoO$_6$ (010), Bi$_2$WO$_6$ (100), Bi$_2$WO$_6$ (010) surfaces have slopes of $\mathscr{S}=$ 0.64, 0.85, 0.87, and 0.85 respectively.}
	\label{fig:Schottky}
\end{figure}
Therefore, the charge-pinning fraction is a critical descriptor of the impact of surface states on the Schottky barrier \cite{Kurtin1969a,Brillson1978,Bard1980}. To determine this important parameter, we calculate the Schottky barrier height of each surface as a function of the difference between the flatband potential and the standard hydrogen evolution potential, as shown in Fig.~\ref{fig:Schottky}. These graphs show a linear trend with a slope of $\sim$0.85-0.87 for Bi$_2$WO$_6$ terminations and the Bi$_2$MoO$_6$ (010) termination. In contrast, the Bi$_2$MoO$_6$ (100) termination has a much lower charge pinning factor, $\sim$0.64. This reflects the wider variance in the charge--voltage curves of Bi$_2$MoO$_6$ (100) surface termination; the adsorption of hydrogen has strong repercussions on the surface dipole and charge distribution for this termination. It should be noted that we do not expect all of the terminations shown here to be stable at the large voltage range shown. These Schottky barriers should be taken as theoretical extrapolations for a metastable phase with a particular termination, allowing us to extract charge--pinning factors. 
\begin{figure*}
	\includegraphics[width=\textwidth]{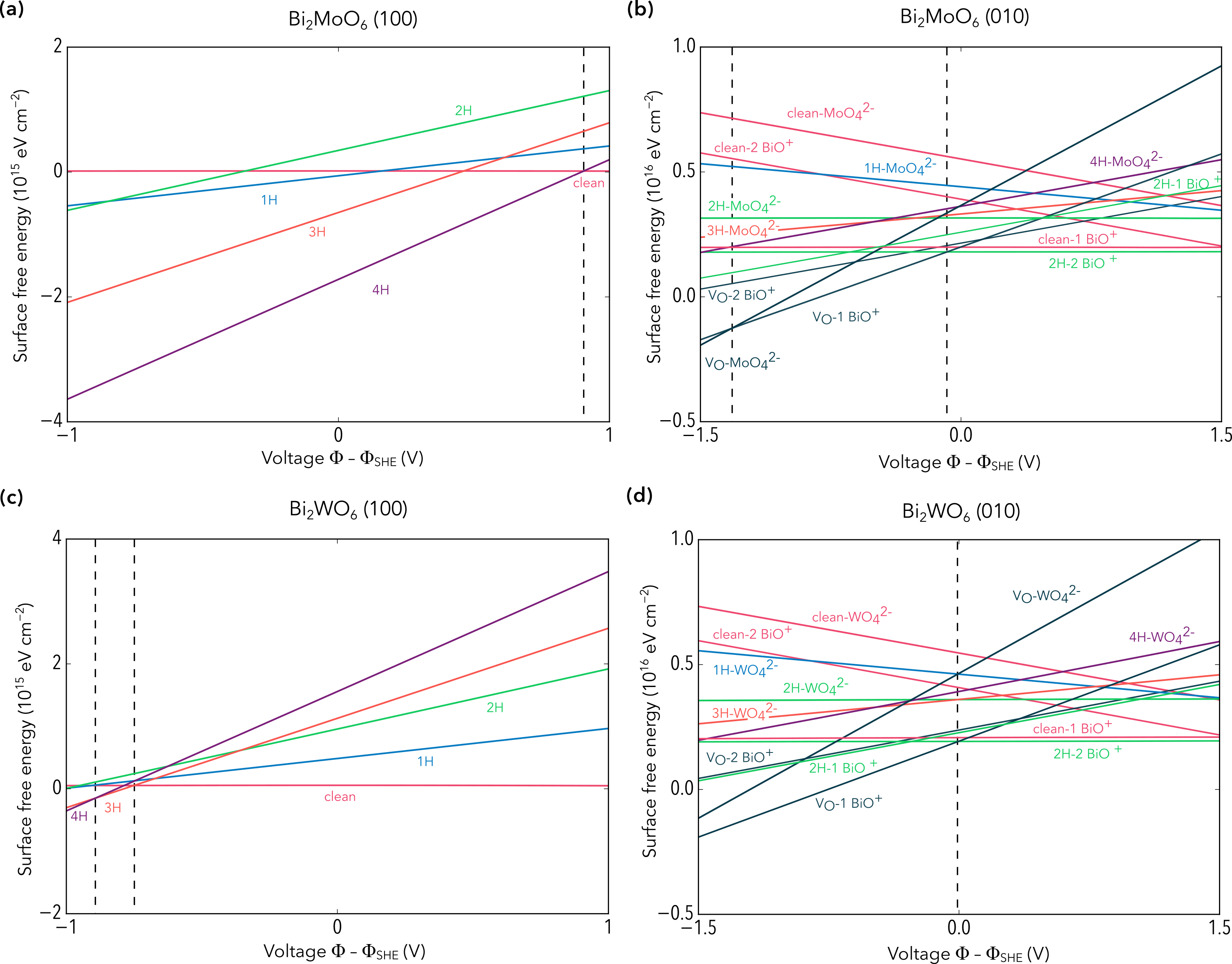}
	\caption{ \small Surface free energy of different adsorbates and layers at the (a) Bi$_2$MoO$_6$ (100), (b) Bi$_2$MoO$_6$ (010), (c) Bi$_2$WO$_6$ (100), (d) Bi$_2$WO$_6$ (010) surfaces, measured at pH=7 under the assumption that the surrounding solution is saturated with BiO$^+$ and W/MoO$_4^{2-}$ ions.}
	\label{fig:stability}
\end{figure*}

Having found the charge--voltage and Schottky barrier relationships, we turn our attention to determining the surface stability of each layer using the techniques outlined above. We calculate the surface stability as a function of potential at a pH of 7 as shown in Fig.~\ref{fig:stability}. Both the Bi$_2$MoO$_6$ and Bi$_2$WO$_6$ (010) surfaces show a surface phase transition from one BiO$^+$ layer at negative potentials to surfaces terminated with two BiO$^+$ ions at higher potentials, with the Bi$_2$MoO$_6$ (010) surface seeing an additional phase transition to MoO$_4^{2-}$ termination at voltages below --1.35 V. Notably, for both the Bi$_2$MoO$_6$ and the Bi$_2$WO$_6$ (010) surface, the unit cell with two BiO$^+$ terminating layers and two hydrogen adsorbed has nearly the same surface free energy as a pristine (``clean'') surface terminated with only one BiO$^+$ layer, making it likely that a mixture between the two different terminations would form in solution. For the Bi$_2$MoO$_6$ (100) surface, the maximally hydrated surface termination tested was the most stable across a broad range of potential only giving way to a pristine interface at $\sim$ 0.9 V. In contrast, for the Bi$_2$WO$_6$ (100) surface, the pristine surface is the most stable across a broad potential range, with hydrated surface terminations only becoming stable at lower voltages.

These electrochemical transitions are particularly important in light of the strong variation of the Schottky barrier height as a function of surface termination. In fact, for the Bi$_2$MoO$_6$ (010) surface, transitioning from a V$_{\rm O}$-MoO$_4^{2-}$ terminated surface to a V$_{\rm O}$-1 BiO$^{+}$ and then to a 2H-2 BiO$^{+}$ surface leads to a change in Schottky barriers from --0.06 V to 0.70 V to --0.18 V. Since the Schottky barrier provides the motive force for charge separation and transfer of electrons from the bulk to the surface of the electrode, the highest magnitude Schottky barrier will see the highest efficiencies. This means the potential window where the V$_{\rm O}$-1 BiO$^{+}$ surface dominates the Bi$_2$MoO$_6$ (010) termination will likely exhibit the most pronounced hydrogen generation. Similarly, the V$_{\rm O}$-1 BiO$^{+}$ surface termination has the highest Schottky barrier of any stable structure for the Bi$_2$MoO$_6$ (010) surface with a Schottky barrier of 0.45 V.  Notably, the high magnitude Schottky barrier of the Bi$_2$WO$_6$ (100) oxygen terminated surface is stable across a much broader range of potentials than for the Bi$_2$WO$_6$ (010) surface, providing a computational interpretation of the experimental conclusions of Saison {\it et al.} \cite{Saison2011a} that the (100) facet is more active than its (010) counterpart. 

\section{Conclusion}
In summary, we calculated the surface structure and electrical characteristics of layered Aurivillius compounds from first principles, addressing the critical problems of determining the charge distribution between the semiconductor and its surface, and of evaluating the surface energy of favored termination. To calculate the equilibrium charge--voltage distribution of the semiconductor--solution, we implemented a Newton--Raphson charge optimization algorithm that has enabled us to effectively compute the interfacial charge distribution as a function of the applied voltage. In addition, to obtain the surface-dependent termination and stability of each layer of a layered material, we calculated the energy of taking individual ionic layers out of solution as a function of potential and pH.

By combining these computational capabilities, we examined the Bi$_2$WO$_6$ and Bi$_2$MoO$_6$ (100) and (010) surfaces, showing a transition from a termination with a single bismuth layer to one with two bismuth layers.  We further demonstrated that oxygen vacancies on a single-bismuth-layer termination gives the highest equilibrium Schottky barriers for both (010) surfaces. Finally, our analysis highlighted that the Bi$_2$WO$_6$ (100) surface has a more favorable Schottky barrier than the (010) surface over a wider potential range, providing electronic-structure evidence for the experimentally observed activity of the (100) surface. Computational studies such as the one presented here offer guidance in optimizing Aurivillius oxides for photocatalytic water splitting. In particular, our study suggested that the Bi$_2$WO$_6$ (100) facet should be a central target for the efficient separation of the photogenerated charge carriers.

\begin{acknowledgements}
	The authors acknowledge primary support from the National Science Foundation under Grant DMR-1654625, and partial support from the 3M Graduate Fellowship and Penn State University Graduate Fellowship. 
\end{acknowledgements}

\end{document}


\title{Supporting Information: ``Voltage-dependent reconstruction of layered Bi$_2$WO$_6$ and Bi$_2$MoO$_6$ photocatalysts and its influence on charge separation for water splitting''}
	
\author{Quinn Campbell}
\author{Daniel Fisher}
\author{Ismaila Dabo}
\affiliation{Department of Materials Science and Engineering, Materials Research Institute, and Penn State Institutes of Energy and the Environment, The Pennsylvania State University, University Park, PA 16802, USA}

\maketitle
\section{Semiconductor--Interface Simulation}

\begin{figure}
	\includegraphics[width=0.7\columnwidth]{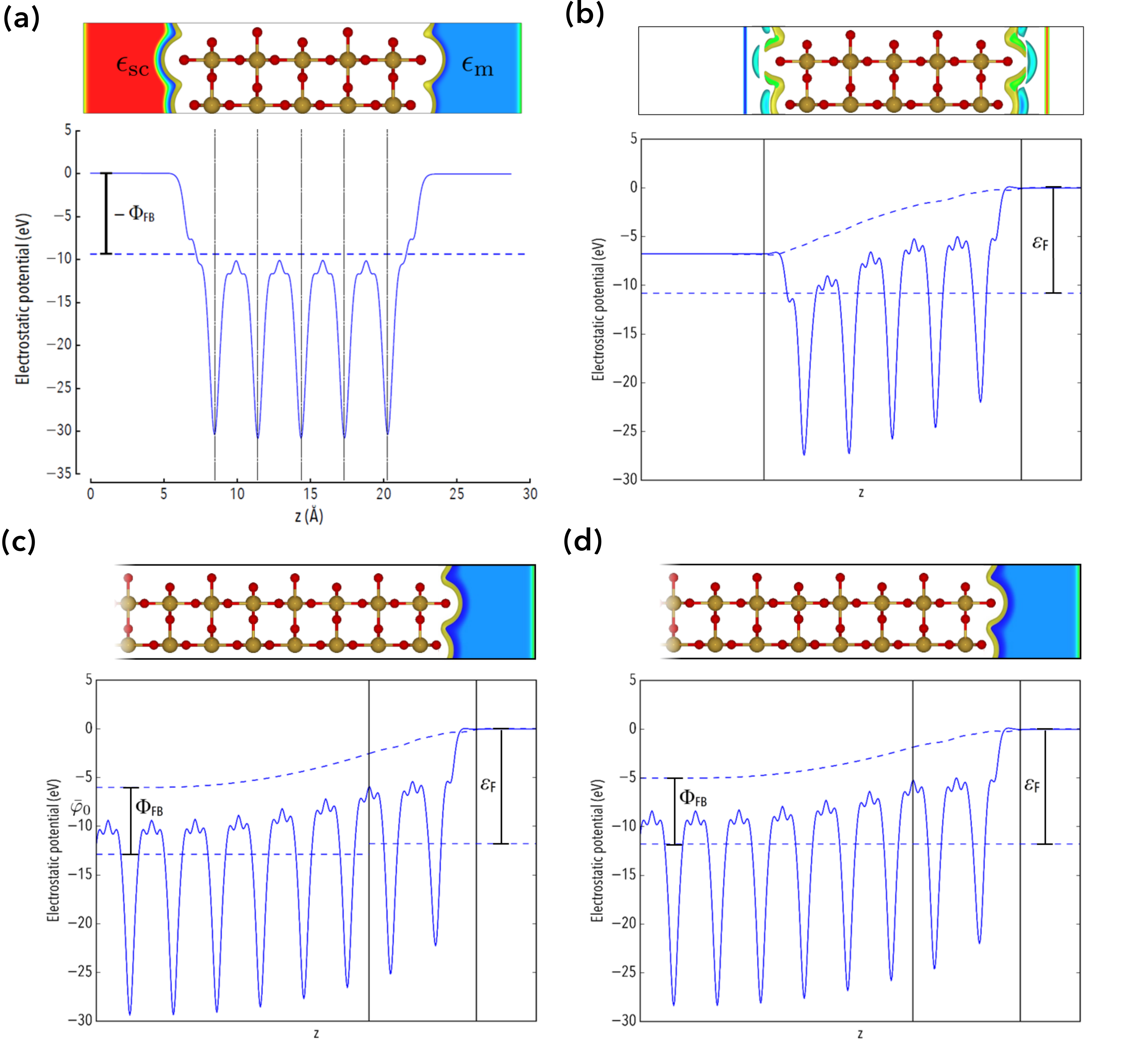}
	\caption{ \small Band bending and alignment of a semiconductor electrode is predicted by initially calculating the potential profile of (a) an electrically neutral slab. To simulate an applied potential (b) Helmholtz planes of countercharge are placed at both surfaces, and charge is added to the slab to maintain charge neutrality. (c) A cutoff place corresponding to the inflection point of the averaged electrostatic potential is added, where the left-hand side is fitted with a Mott--Schottky extrapolation. (d) To find the equilibrium charge distribution, the Fermi levels of the bulk semiconductor $\Phi_{\rm FB}$ and the interface $\epsilon_{\rm F}$ are matched. Fig adopted from Ref.~\onlinecite{Campbell2017}.}
	\label{fig:method}
\end{figure}

The calculation of the equilibrium charge--voltage distribution of a semiconductor--solution system begins with a simulation of an electrically neutral system as shown in Fig.~\ref{fig:method}a. Using the SCCS model, we apply a different dielectric constant to each side of the slab, representing the dielectric constant of the semiconductor $\epsilon_{\rm sc}$ and the dielectric constant of the surrounding medium $\epsilon_{\rm m}$, in the case of water $\epsilon_{\rm m} = 78.3$. When the potential within the solution region is aligned to zero, the flatband potential $\Phi_{\rm FB}$ can then be taken as the opposite of the Fermi level.

To predict the electrification of the electrode we then place planar countercharges on either side of the slab, as shown in Fig.~\ref{fig:method}b. The electrode is assigned a total charge of $q$, split between the quantum-mechanical region $q^{\rm surf}$ and the bulk semiconductor $q^{\rm bulk}$ such that $q=q^{\rm bulk}+q^{\rm surf}$. Here, the charge $q^{\rm bulk}$ is placed on the counter charge on the semiconductor side. We then assign the opposite countercharge plane as $-q$, ensuring that the supercell is charge neutral. The plane of countercharge within the solution and the SCCS model describe a Helmholtz model of the electrode solution interaction well; however, a Mott--Schottky potential distribution needs to be applied to accurately describe the bulk semiconductor.

The Mott--Schottky distribution can be found based on the derivative of the macroscopic potential $\bar \Phi$. Here, we determine $\bar \Phi$ by calculating the difference between the macroscopic average of the electrostatic potential for the charged and neutral slab. To avoid spurious surface interactions, we choose a cutoff value in the z direction $z_{\rm c}$, placed at the inflection point of an electrode's macroscopic potential, as shown in Fig.~\ref{fig:method}c.  On one side of the cutoff, the electrode surface will be described using the quantum-mechanical calculations. To the other side of the cutoff, the electrode will behave as a bulk semiconductor following the Mott--Schottky equations. Using this knowledge, we can then calculate the bulk potential of the semiconductor as $\bar \Phi_0 = \bar {\Phi}(z_{\rm c}) -  k_{\rm B}T - \frac{\epsilon_0 }{2 \mathscr N} (\frac{d \bar \Phi}{dz}(z_{\rm c}))^2$ for an $n$-type semiconductor. Here, $\mathscr N$ is the dopant concentration of the semiconductor electrode, $k_{\rm B}$ is the Boltzmann constant, and $T$ is the ambient temperature. From this bulk potential, we can find the Fermi level of the bulk semiconductor by using the charge--neutral Fermi level of the slab $E_{\rm F}^{\circ}$ and adding it to the potential of the bulk semiconductor found earlier ($E_{\rm F}^{\rm bulk} = \bar \Phi_0 + E_{\rm F}^{\circ}$). 

To find the equilibrium charge distribution of the semiconductor, we need to find the charge distribution where the Fermi level of the bulk semiconductor and the surface are in equilibrium as shown in Fig.~\ref{fig:method}d. This enables us to find the equilibrium charge--voltage response of the system. In this work, we develop an optimization algorithm allowing us to automatically find the equilibrium distribution.

\section{Convergence of surface energy with respect to number of layers}

\begin{figure}
	\includegraphics[width=0.7\columnwidth]{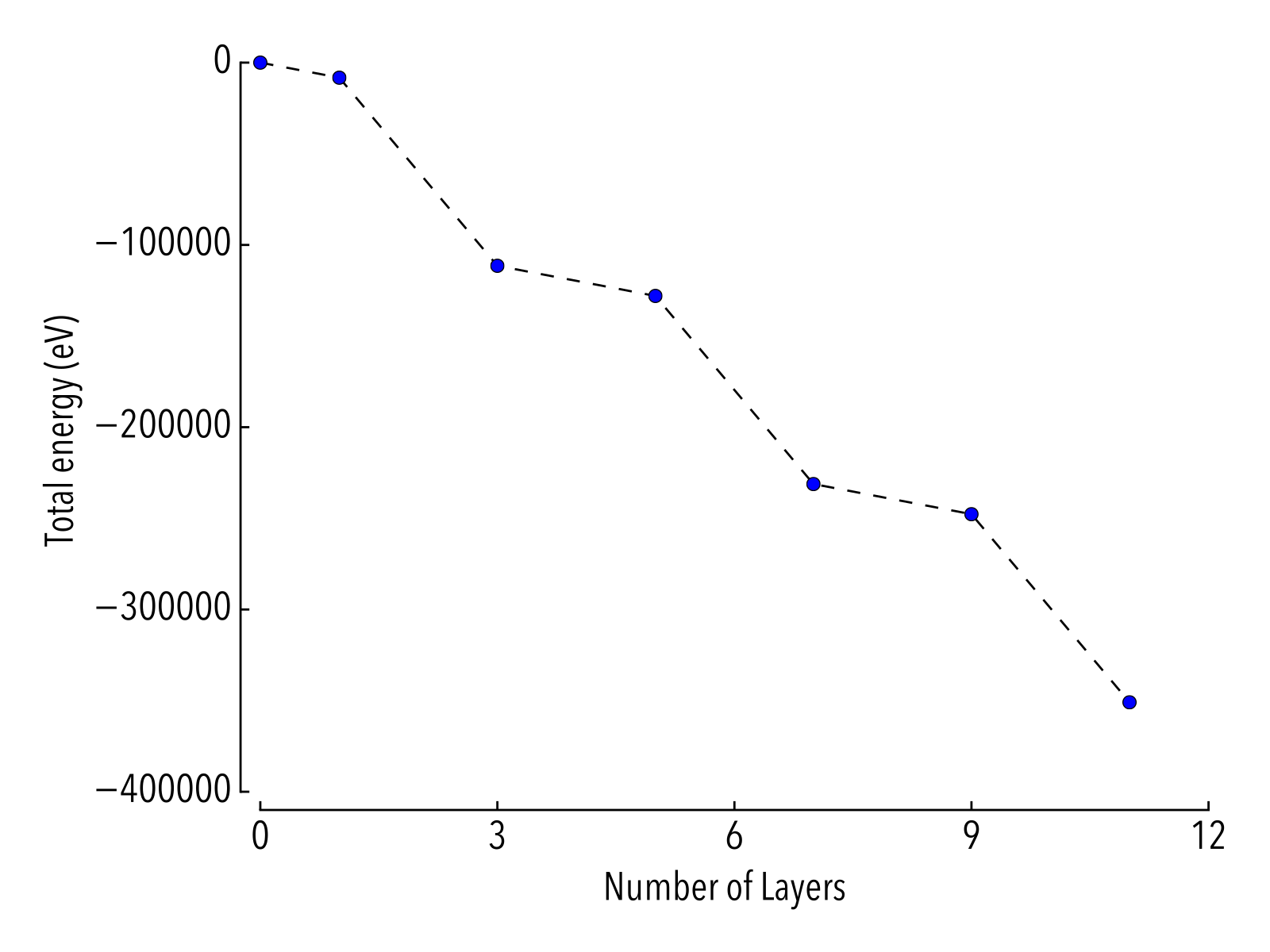}
	\caption{ \small Total calculated energy of slabs as a function of number of layers simulated.}
	\label{fig:surf-e}
\end{figure}

When running slab calculations, it is important to make sure that the slabs be of sufficient thickness such that surface energies of terminal layers are converged. In Fig.~\ref{fig:surf-e}, we show the calculated total energy of the Bi$_2$MoO$_6$ slab terminated in the (010) direction. It is clear that the energy does not follow the linear pattern typical of most slab calculations. This is because Aurivillius oxides alternate between adding MoO$_4^{2-}$ layers and 2BiO$^+$ layers. Thus, we must ensure instead that the slope between corresponding layers stays constant. Using this criterion and the results of Table \ref{Table_1}, we determine five layer slabs to be the best compromise between computational cost and accuracy.

 \begin{table}
 	\small
 	\centering
 	\caption{The energy of adding one additional layer to the Bi$_2$MoO$_6$ surface.}
 	\begin{tabular*}{0.8\columnwidth}{l @{\extracolsep{\fill}} ccc}		     \\
 		\hline \hline
 		Layers &   Added layer & Total energy (eV) & Energy of one additional layer (eV) \\ 
 		\hline \\
 		0  & &0&  \\
 		1 &  MoO$_4^{2-}$&-8264.82 & -8264.82 \\
 		3 & 2BiO${+}$& -111454.26& -51594.7   \\
 		5 & MoO$_4^{2-}$& -127992.83& -8269.29\\
 		7 & 2BiO${+}$& -231185.91& -51596.5  \\
 		9  & MoO$_4^{2-}$&-247724.42 & -8269.25\\
 		11  & 2BiO${+}$& -350917.33  & -51596.5\\	
 		
 		\\ \hline \hline
 	\end{tabular*}%
 	\label{Table_1}%
 \end{table}

